\documentclass[preprint2]{aastex}

\newcommand{\mujy}{$\mu$Jy\ }
\newcommand{\sigWSRT}{\sigma_{\tiny \mbox{WSRT}}}
\newcommand{\sigNVSS}{\sigma_{\tiny \mbox{NVSS}}}
\newcommand{\SNRWSRT}{\mbox{SNR}_{\tiny \mbox{WSRT}}}
\newcommand{\SNRNVSS}{\mbox{SNR}_{\tiny \mbox{NVSS}}}
\newcommand{\sWSRT}{S_{\tiny \mbox{WSRT} }}
\newcommand{\sNVSS}{S_{\tiny \mbox{NVSS} }}
\newcommand{\sTRUE}{S_{\tiny \mbox{true} }}

\newcommand{\sSNR}{{\tiny \mbox{SNR} }}
\newcommand{\NVSS}{${\small \mbox{NVSS}}$}
\newcommand{\GB}{${\small \mbox{87GB}}$}
\newcommand{\WSRT}{${\small \mbox{WSRT}}$}
\newcommand{\APM}{${\small \mbox{APM}}$}
\newcommand{\FIRST}{${\small \mbox{FIRST}}$}
\newcommand{\WENSS}{${\small \mbox{WENSS}}$}
\newcommand{\VLA}{${\small \mbox{VLA}}$}
\newcommand{\NOAO}{${\small \mbox{NOAO}}$}
\newcommand{\NRAO}{${\small \mbox{NRAO}}$}
\newcommand{\MERLIN}{${\small \mbox{MERLIN}}$}
\newcommand{\VLT}{${\small \mbox{VLT}}$}
\newcommand{\IRS}{${\small \mbox{IRS}}$}
\newcommand{\IRAC}{${\small \mbox{IRAC}}$}
\newcommand{\MIPS}{${\small \mbox{MIPS}}$}
\newcommand{\SIRTF}{${\small \mbox{SIRTF}}$}
\newcommand{\DCB}{${\small \mbox{DCB}}$}
\newcommand{\SNR}{${\small \mbox{SNR}}$}
\newcommand{\AGN}{${\small \mbox{AGN}}$}
\newcommand{\FWHM}{${\small \mbox{FWHM}}$}

\shorttitle{WSRT Bootes Deep Field}

\begin{document}
\title{Deep Westerbork 1.4 GHz Imaging of the Bootes Field}

\author{W. H. de Vries\altaffilmark{1}, R. Morganti\altaffilmark{2},
H. J. A. R\"ottgering\altaffilmark{3}, R. Vermeulen\altaffilmark{2},
W. van Breugel\altaffilmark{1}, R. Rengelink\altaffilmark{3}, \&
M. J. Jarvis\altaffilmark{3}}

\altaffiltext{1}{LLNL / IGPP, 7000 East Ave, L-413, Livermore, CA
94550 (devries1@llnl.gov, vanbreugel1@llnl.gov)}
\altaffiltext{2}{ASTRON, P.O. Box 2, 7990 AA, Dwingeloo, The
Netherlands (morganti@nfra.nl, rvermeulen@nfra.nl)}
\altaffiltext{3}{Leiden Observatory, P.O. Box 9513, 2300 RA, Leiden,
The Netherlands (rottgeri@strw.leidenuniv.nl,
rengelin@strw.leidenuniv.nl, jarvis@strw.leidenuniv.nl)}

\begin{abstract}

We present the results from our deep (16$\times$12 hour) Westerbork
Synthesis Radio Telescope (\WSRT) observations of the approximately 7
square degree Bootes Deep Field, centered at $14^{\rm h}32^{\rm
m}05\fs75$, $34\arcdeg16\arcmin47\farcs5$ (J2000). Our survey consists
of 42 discrete pointings, with enough overlap to ensure a uniform
sensitivity across the entire field, with a limiting sensitivity of
28\mujy (1$\sigma_{\mbox{rms}}$). The catalog contains 3172 distinct
sources, of which 316 are resolved by the $13\arcsec\times27\arcsec$
beam.  The Bootes field is part of the optical/near-infrared imaging
and spectroscopy survey effort conducted at various institutions. The
combination of these data sets, and the deep nature of the radio
observations will allow unique studies of a large range of topics
including the redshift evolution of the luminosity function of radio
sources, the K-$z$ relation and the clustering environment of radio
galaxies, the radio / far-infrared correlation for distant starbursts,
and the nature of obscured radio loud \AGN.

\end{abstract}


\section{Introduction}

One of the main goals of radio astronomy is to fully understand the
physics of the population of extragalactic radio sources (RSs). Issues
include the onset and demise of the radio activity and related
starbursts, the influence of the environment on the characteristics of
the RSs and the appearance of the first RSs and their relation to the
formation of galaxies, massive black holes and the reionization of the
universe.

Detailed investigation of complete samples of bright RSs with redshift
information have been carried out over the last decades (e.g., 3CRR:
Laing, Riley \& Longair 1983; 6CE: Eales 1985, Eales et al. 1997,
Rawlings, Eales \& Lacy 2001) and led to many interesting
discoveries. For example, it is now well established that the comoving
number density of $z \approx 1$ powerful RSs is about two orders of
magnitude larger than it is locally (e.g., Longair 1966, Dunlop \&
Peacock 1990). Another example is that the environment of RSs changes
with redshift, with bright RSs at higher redshifts located in denser
environments than locally (e.g., Best, Longair \& R\"ottgering 1998).

Through the selection of RSs that are bright and have very steep radio
spectra ($<-1.3$), more than 150 powerful galaxies with $2 < z < 5.2$
have been found \citep{debreuck00}. The large starformation rates
\citep{dey97} and extremely clumpy optical/IR morphologies
\citep{pentericci99} provide strong evidence that these galaxies are
massive galaxies close to the epoch of formation. Powerful radio
emission is most likely caused by accretion onto massive black holes
(M$_{\rm BH} > 10^9$ M$_\sun$; McLure et al. 1999, Laor 2000),
indicating that such massive black holes formed alongside or possibly
before the formation epoch of their host galaxies (e.g., Kauffmann \&
Haehnelt 2000). Recent \VLT\ observations have revealed the existence
of a large scale structure of Ly-$\alpha$ emitting galaxies around the
radio galaxy 1138$-$262 ($z=2.2$), reinforcing the idea that high
redshift radio galaxies (HzRGs) can be used as tracers of
proto-clusters of galaxies \citep{pentericci00}.

At very faint flux densities (e.g., a few tens of \mujy at 1.4 GHz)
the radio source counts are dominated by the $z\la1$ starburst
population (e.g., Richards et al. 1999). Long observations with the
\VLA\ + \MERLIN\ \citep{richards00, muxlow99} and \WSRT\
\citep{garrett00} reach such faint levels, and have enabled important
constraints to be placed upon the redshifts and nature of distant
starburst galaxies.

\subsection{Survey Rationale}

\begin{figure}[tb]
\plotone{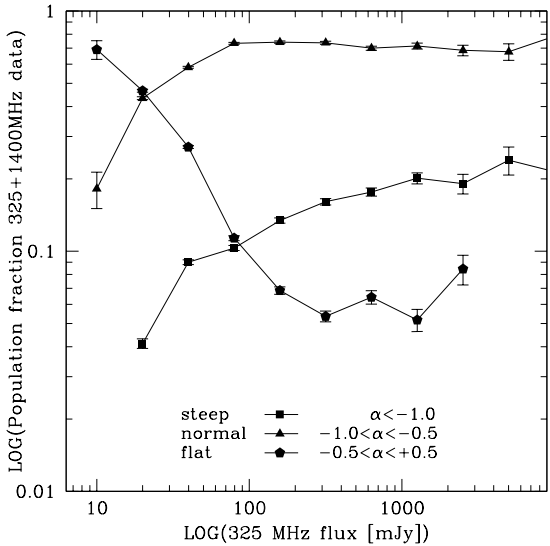}
\caption{Composition of the radio source population as function of 325
MHz flux. The fraction of flat spectrum sources increases dramatically
with decreasing flux, whereas the steep spectrum component (dominated
by \AGN) drops precipitously. Data are from the source overlap between
the \WENSS\ and \NVSS\ surveys, at 325 and 1400 MHz respectively. The
errorbars represent Poissonian errors, and may be smaller than the
symbol size in some cases. Note that \WENSS\ sample incompleteness
sets in around 20 mJy, affecting the steep source count most.}
\label{specPop}
\end{figure}

Our \WSRT\ survey reaches a 1$\sigma$ detection threshold of 28 \mujy
at 1.4 GHz, within a factor of 2-3 of the deepest radio observations
carried out so far (cf. Windhorst et al. 1999, Richards 2000). The
surveyed area is, however, large enough (about 7 square degrees) to
yield enough sources to not be severely affected by low number
statistics for the RS populations under scrutiny. As can be seen in
Fig.~\ref{specPop}, the composition of the radio population changes
dramatically towards lower flux density limits. The higher flux levels
are dominated by RSs with steep spectra ($< -0.5$), and a cross-over
to flat-spectrum sources appears to occur at the 10-100 mJy (at 325
MHz) level. The former are mostly identified with (powerful) radio
sources residing in massive ellipticals (e.g., Eales et al. 1997),
whereas the latter can be tied to a population of starforming
late-type galaxies, especially towards sub-mJy 1.4 GHz flux density
levels (e.g., Windhorst 1999, Richards et al. 1999). Therefore, deeper
observations will not only increase the number of detected sources,
but it will also provide a better handle on the relative makeup of the
radio population at \mujy\ flux density levels.
 
\begin{figure}[tb]
\plotone{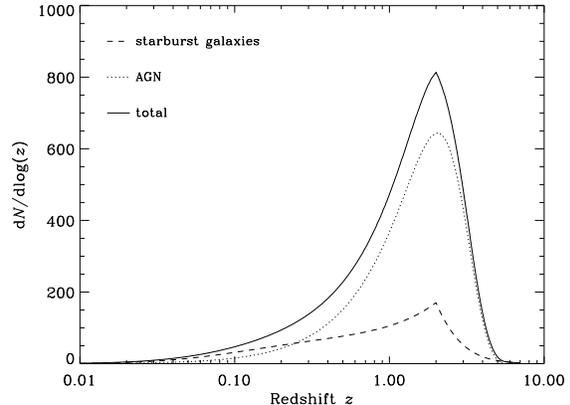}
\caption{Expected redshift distribution of the radio source population
for our survey, based on the luminosity functions of Dunlop \& Peacock
(1990, PLE model), and Hopkins et al. (1998, the Phoenix survey).}
\label{zdistPred}
\end{figure}

The radio source population properties do not only change as a
function of flux density, they also vary with redshift. Using radio
luminosity functions for AGN \citep{dunlop90} and starbursting
populations \citep{hopkins98}, we can model the expected source counts
for a given redshift (and limiting flux density). This is plotted in
Fig.~\ref{zdistPred} for a limiting flux density of 140$\mu$Jy. Beyond
a redshift of $z\sim0.3$ the number counts for our survey are expected
to be dominated by the AGN population. The combination of information
presented in Figs.~\ref{specPop} and \ref{zdistPred} makes it clear
how dependent on limiting flux density a perceived radio source
population is. Indeed, for a limiting flux density of 1 mJy (at 1.4
GHz) the population is dominated by AGN-type sources at all redshifts
\citep{hopkins98}.

The survey can detect radio sources at the FR~I / FR~II break level
(10$^{25}$ W Hz$^{-1}$ at 1400 MHz, Owen \& White 1991) out to a
redshift\footnote{We adopt H$_{\rm o}$=65 km s$^{-1}$ Mpc$^{-1}$,
$\Omega_{\rm M} = 0.3$, and $\Omega_\Lambda = 0.7$ throughout this
paper.} of $z\sim3$. Fainter FR~I type sources with a mean radio power
of 10$^{24}$ W Hz$^{-1}$ drop out of the sample around $z\approx1$,
and the much fainter starforming systems will not be detected beyond
$z\sim0.15$ at the 10$^{22}$ W Hz$^{-1}$ level. The local starbursting
system M82 has, for comparison, a radio spectral power of $10^{21.99}$
W Hz$^{-1}$, given its 1400 MHz flux density of 8.363 Jy
\citep{white92} and the adopted cosmology.  Thus the survey provides
ample data for evolutionary studies of {\em radio loud} systems out to
at least a redshift of 1, whereas a more complete census of radio
sources down to the spiral / starburst level has to be limited to
sources within $z\sim0.1$.

\subsection{Relation to non-Radio Surveys}

Key ingredients for follow-up studies are optical / near-infrared
identifications and redshift information for at least a large fraction
of the radio sources. For this purpose a number of other surveys are
either being carried out or are planned for the same part of the
sky. These include:

{\it The \NOAO\ Deep Wide-Field Survey} (PI's Jannuzi and Dey). This
survey consists of a Northern and a Southern part, with the former
field located in Bootes (near the North Galactic Pole), covering a
$3\times3$ degree region, and the latter located in a $2\times4.5$
degree equatorial region in Cetus. Both fields have been selected for
their low mid- to far-infrared cirrus emission, their low H~{\small I}
column densities and the availability of high resolution
($\sim5\arcsec$) \VLA\ -- \FIRST\ survey radio data. Of particular
interest to our program is the Bootes field, which will be imaged to a
limiting surface brightness of about 28$^{\rm th}$ magnitude per
square arcsecond in $B$, $V$, and $R$, and down to $\sim22^{\rm nd}$
magnitude per square arcsec in $J$, $H$, and $K$. These detection
limits will permit the optical and near-infrared study of faint,
{sub-L$_{*}$} galaxies out to redshifts of about unity (an L$_{*}$
galaxy will have $K=19.82$ at $z=1$, based on an absolute magnitude of
$K=-24.44\pm0.06$, cf. Kochanek et al. 2001).  The typical host
galaxies of luminous radio sources, with masses well in excess of
L$_{*}$ galaxies, can be detected out to very large redshifts (based
on the $K-z$ diagram, e.g., Jarvis et al. 2001, De Breuck et
al. 2002). Given our radio source population mix of powerful radio
sources associated with intrinsically bright galaxies at high redshift
and less luminous starforming systems at much lower redshifts, we
expect to be able to detect optical / near-infrared counterparts for
most of them. The \NOAO\ survey limits are well matched to our
expected counterpart population.

{\it The \IRAC\ Shallow Survey} (PI. Eisenhardt). The Bootes field will
be covered by \SIRTF's InfraRed Array Camera (\IRAC) in four IR bands
ranging from 3.6$\mu$m to 8$\mu$m. Coverage towards the longer IR
bands up to 160$\mu$m will be provided by the Multiband Imaging
Photometer (\MIPS), and some spectroscopy by the InfraRed Spectrograph
(\IRS, PI in both cases is J. Houck).

The \NOAO\ and \SIRTF\ wide field surveys are aimed to study, among
other things, (i) the evolution of large-scale structure from $z\sim
1-4$, (ii) the formation and evolution of ellipticals and starforming
galaxies, and (iii) the detection of very distant ($z>4$) young
galaxies and quasars. The \SIRTF\ IRAC and MIPS observations will also
detect star-forming galaxies at mid- to far-infrared wavelengths. It
is this multi-wavelength aspect of the project, covering a large
fraction of the electromagnetic spectrum (2 radio frequencies, several
optical and near-infrared bands, and the mid- to far-infrared space
based \SIRTF\ observations) that distinguishes this effort from other
deep radio-optical/near-infrared surveys like the Phoenix survey,
\citep{hopkins98,georgakakis99}, and the Australia Telescope ESO Slice
Project (ATESP) \citep{prandoni00a,prandoni00b,prandoni01}.

\subsection{Relation to other Radio Surveys}

\begin{figure}[tb]
\plotone{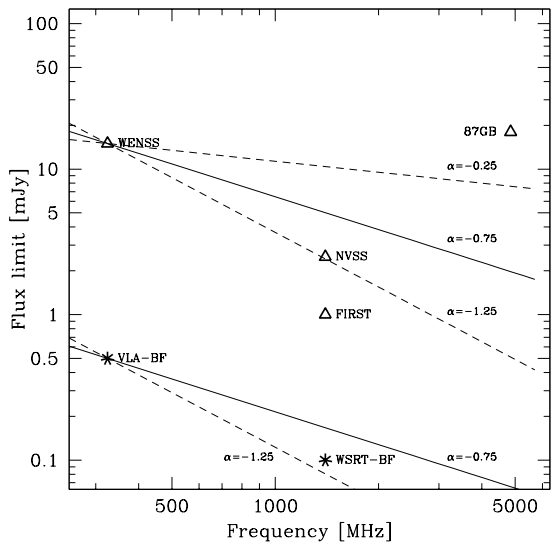}
\caption{Comparison between the various radio surveys covering the
Bootes field. The triangles represent literature surveys, and the
stars are our Bootes \WSRT\ and \VLA\ surveys. Overplotted are
representative radio spectra with varying spectral indices:
$\alpha=-0.25, -0.75, \mbox{and} -1.25$, for flat-, normal-, and
steep-spectrum radio sources. Any given radio source in the \VLA-BF
survey with a slopes shallower than $-1.25$ should also be present in
\WSRT-BF. On the other hand, of the sources in \WSRT-BF fainter than
the \VLA-BF limit ($\sim0.5$ mJy), only the objects with slopes steeper
than $-1.25$ should be present in \VLA-BF.}
\label{surveyComp}
\end{figure}

The Bootes field has been covered by previous radio surveys, most
notably by the Westerbork Northern Sky Survey (\WENSS, Rengelink et
al. 1997) at 325 MHz, by the \NRAO\ \VLA\ Sky Survey (\NVSS, Condon et
al. 1998), and by the Faint Images of The Radio Sky at Twenty-cm
survey (\FIRST, Becker et al. 1995), both at 1.4 GHz. A comparison
between the literature surveys and our Bootes surveys (\WSRT\ at 1.4
GHz in this paper, and \VLA\ 325 MHz) is shown in
Fig.~\ref{surveyComp}, and tabulated in Table~1. The varying survey
depths and frequencies make combining catalogs to obtain spectral
index information less than straightforward. For instance, combining
\NVSS\ and \GB\ (at 5 GHz, Gregory \& Condon 1991) only makes sense if
one is interested in radio sources with strongly inverted spectra. For
our purpose, since the bulk of the radio source population has a
spectral index of around $-0.75$ (cf. Fig.~\ref{spixDist}), a
combination of surveys like the \WENSS\ and \NVSS\ / \FIRST\ is best,
as can be inferred from the overplotted common radio spectra in
Fig.~\ref{surveyComp}.

However, these surveys do not go deep enough to effectively probe the
transition in radio source population occurring around the 1 mJy level
(cf. Figs.~\ref{specPop}, \ref{dnds}, and, e.g., Windhorst et
al. 1999). Our \WSRT\ observations do go deep enough, but will need
low frequency data of matching sensitivity. We use the \VLA\ at 325
MHz for this purpose, and the data from this program will be described
in an subsequent paper. However, both the \NVSS\ and \FIRST\ survey
data have been used to calibrate our survey flux densities and
positions (cf. Sections~\ref{fluxacc} and \ref{posacc}).
  
\section{Observations}

The observations were carried out by the Westerbork Synthesis Radio
Telescope (\WSRT) operating at 1.380 GHz. The \WSRT\ consists of 14
25m telescopes, arranged in a 2.7km East-West configuration. As the
back-end, we used the Digital Continuum Backend (\DCB) with 8
sub-bands of 10 MHz bandwidth each. The smallest baseline (9$-$A) was
set to 54m to limit shadowing at the expense of a reduction in large
spatial structure sensitivity ($\sim800\arcsec$ for this minimum
baseline and frequency).

\subsection{Field Layout and Instrumental Setup}

We designed a survey layout consisting of 42 discrete pointings. The
separation between the grid points was chosen to be 60\%\ of the
\FWHM\ of the primary beam. Given the used tiling and the known
attenuation of the beam, a more or less uniform noise-background is
obtained with this spacing (85\%\ of the survey area has a local rms
within $\sqrt{2}\sigma_{\mbox{\tiny median}}$,
cf. Sect~\ref{complreliab} and Fig.~\ref{noiseMap}). This strategy is
similar to the one used for the Australia Telescope Compact Array
(ATCA) ATESP survey (see Prandoni et al. 2000a for a detailed
description).  The total number of pointings (42) was dictated by the
need to cover the 325 MHz \VLA\ primary beam with a uniform
sensitivity (noise) level.

During each observing block of 12 hours, the telescopes were
continuously cycled between 3 individual grid positions. This
``mosaicing'' mode \citep{kolkman93} allows multiple fields to be
observed, while still retaining a 12 hour $uv$-coverage for the fields
individually (albeit sampled non-continuously). The basic integration
time for the observations is 10 seconds. A typical observing cycle can
be broken down to a 10s slew-time between the grid positions, and
$5\times10$s on-source time. Even though the net slewing time is less
than 10s, some extra time is needed for the array to settle itself
after the move. The observing efficiency with this scheme is therefore
83.3\%\, resulting in $3 \times 200^{\rm m}$ net observing time per 12
hour cycle. Given the total allocated time for this project (192
hours), we used the remaining two 12 hour blocks to cycle through {\em
all} 42 positions. In this setup, each position was revisited every 42
minutes (instead of every 3 minutes), resulting in a rather sparse
$uv$-sampling and less than $15^{\rm m}$ on-source time per block.
    
Each 12 hour block was sandwiched between two phase and polarization
calibrators, typically 3C~286 and 3C~147, more than adequate given the
system stability. A log of the observations can be obtained from
the ftp-site (cf. Sect~\ref{ftpsite}).

\section{Reduction}

The mosaic was reduced, calibrated, and assembled using the MIRIAD
\citep{sault95} software package. The data was typically of high
quality, and only a few percent had to be flagged. Usually the bad
data was concentrated in channel 5 (of the 8), which, around 1380 MHz,
is the frequency band most affected by interference due to the Global
Positioning System.  Every field pointing was mapped using a multi
frequency synthesis approach, where the measurements of the 8 bands
individually are gridded simultaneously in the uv plane. This
significantly reduces bandwidth smearing problems. Then a three step
iterative phase self-calibration cycle was used, using typically
around 100\,000 clean iterations. The clean was done down to the 3
sigma level, so as not to incorporate too much flux in faint sources
that does not belong there. In the few fields with strong sources
present we performed amplitude self calibration as well, in all other
cases this did not improve the final outcome. In the fourth, and final
cycle, spectral index effects on beam-shape were taken into account
\citep{sault99}.  The final maps improved significantly by correcting
for the small primary beam shape variations across the 8 10 MHz wide
frequency channels.

After all the field pointings were reduced in this manner, they were
assembled into a final mosaic. This step basically adds up the maps
after performing the proper primary beam correction. Since the dirty
beam changes slightly across the field, we restored all the data with
a fixed synthesized beam of $13.0\arcsec \times 27.0\arcsec$, at a
position angle of zero degrees. The mosaic was then further mapped
onto $4\arcsec\times4\arcsec$ pixels, for a total of $\sim2750^2$ pixels.

\section{Results} \label{ftpsite}

We used automated routines for the source extraction and catalog
creation. These were slightly modified from their \WENSS\
counterparts, but the applied methods are exactly the same, all of
which are described in detail in \citet{rengelink97}.

The software works on rectangular patches of sky only, so we tiled the
circular overall shape of the survey into three rectangular areas, as
outlined in Fig.~\ref{noiseMap}. All of the low noise areas have been
included this way, and only a few parts of the noisy edges have not
been cataloged. The total cataloged survey area covers 6.68 square
degrees.

Part of the catalog, to illustrate its format, has been listed in
Table~2. The full version (with 3172 sources) can be obtained through
anonymous ftp to ftp://ftp.nfra.nl/pub/Bootes. The complete mosaic,
individual pointing maps, and tables with various additional data are
available from the same address, all of which are described in the
README file.
    
\subsection{Flux Accuracy and Error Estimates} \label{fluxacc}

\begin{figure}[tb]
\plotone{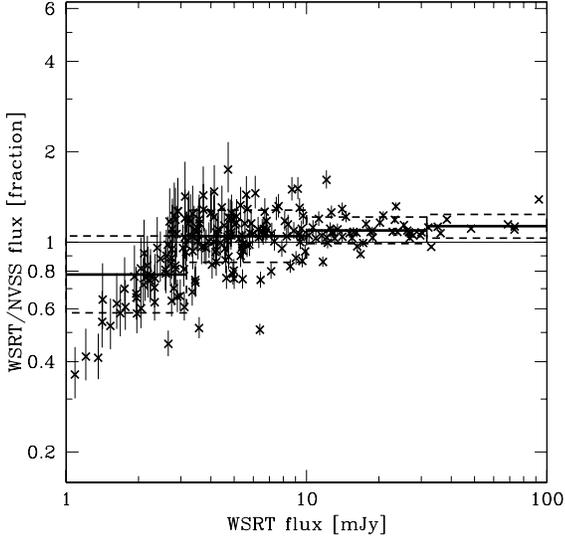}
\caption{Flux density comparison between sources in common to \NVSS\
and \WSRT.  Only unresolved \WSRT\ sources (and hence unresolved in
\NVSS) have been included. The increase in \NVSS\ flux density
relative to the \WSRT\ close to its detection limit is possibly due a
combination of Malmquist and \NVSS\ clean biases. Errorbars are
1$\sigma$ errors, and may be smaller than the symbol size in some
cases.}
\label{fluxComp}
\end{figure}

\begin{figure}[tb]
\plotone{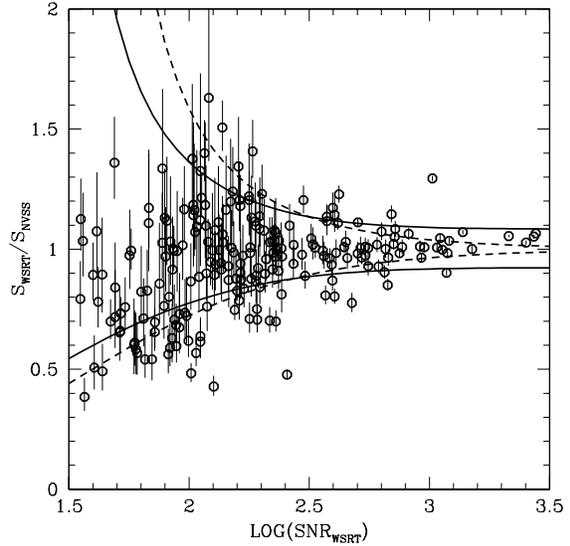}
\caption{Relative flux density errors, as function of Signal-to-Noise
ratio. \WSRT\ fluxes are compared to \NVSS\ fluxes for unresolved
radio sources present in both catalogs. The overplotted solid lines
are expected upper and lower envelopes to the flux density fraction,
assuming values for $C_1$ and $C_2$ (cf. Eqn.~\ref{modelq3}) of $0.04$
and $1.3$ respectively. The dashed lines are for $C_1=0.0$ and
$C_2=2.0$, and illustrate that $C_1$ is most dominant at high \SNR\,
while $C_2$ is it at low \SNR.}
\label{fluxSigma}
\end{figure}

We have compared the flux densities of unresolved sources present in
both our uncalibrated Bootes and the \NVSS\ catalog. Since the
resolution of \NVSS\ (at 45\arcsec) is slightly worse than ours
($13\arcsec\times27\arcsec$), a source which is unresolved in our
catalog is consequently unresolved in \NVSS. While we could compare
our fluxes to the deeper \FIRST\ data, the latter's much higher
angular resolution typically resolves point sources in our catalog,
making a direct comparison difficult. The results of the comparison
are plotted in Fig.~\ref{fluxComp}. It is clear from the plot that our
uncalibrated fluxes are a little too high in comparison to the \NVSS\
fluxes, at least for $S > 10$ mJy. Below these fluxes, the \NVSS\
values are systematically too high, and are presumably due to a
combination of Malmquist and clean biases. This overestimate of \NVSS\
fluxes close to their detection limit ($\sim2.5$ mJy) is also evident
in the Condon et al. (1998) comparison of \NVSS\ fluxes to deep \WSRT\
\citep{katgert85} flux densities, cf. Fig.~31 in \citet{condon98}.
Also, \citet{prandoni00b} noticed the same effect in comparing their
ATCA radio survey fluxes to the \NVSS\ values.

Using flux density weighting, we calculated the offset to be
$4.5\pm3.0$\%\ too high. We reduced our fluxes accordingly
(cf. Fig.~\ref{fluxSigma}). Following \citet{rengelink97}, the
relative flux density errors can be written as:

\begin{equation}
\frac{\sigma_S}{S} = \left( C_1^2 + C_2^2 \left( \frac{\sigma_{\mbox{rms}}}{S} 
\right)^2 \right)^{\frac{1}{2}}
\label{modelq1}
\end{equation}

\noindent This equation reflects the two components of the measurement
error, with $C_1$ due to a constant systematic error, and $C_2$ being
dependent on the Signal-to-Noise Ratio (\SNR). Ideally, one would like
to compare our measured (and corrected) fluxes to their true values in
order to determine the constants $C_1$ and $C_2$ which best fit the
observed $\sWSRT / \sTRUE $ ratio. Unfortunately, we do not have such
a control data set, instead we use the \NVSS\ measurements. If we
assume that the \NVSS\ measurements have a similar error dependence,
we can define the flux density ratio as:

\begin{equation}
\frac{\sWSRT}{\sNVSS} = \frac
    {1 \pm \left( C_1^2 + C_2^2\left( \frac{\sigWSRT}{\sTRUE} \right)^2\right)^\frac{1}{2}}
    {1 \pm \left( C_1^2 + C_2^2\left( \frac{\sigNVSS}{\sTRUE} \right)^2\right)^\frac{1}{2}}
\label{modelq2}
\end{equation}

\noindent with $\sigWSRT$ and $\sigNVSS$ being the median noise in the
sky, and $\sTRUE$ the true value of the source flux. The $\sigma$'s
are quoted as $0.45$ mJy for the \NVSS\ \citep{condon98}, and
$0.028$ mJy for our survey. If we further assume that $\sWSRT$ and
$\sNVSS$ are approximately equal to $\sTRUE$, and that $\mbox{SNR}_X =
S_X / \sigma_X$ with $X$ being either \WSRT\ or \NVSS, we can rewrite
Eqn.~\ref{modelq2} as:

\begin{equation}
\frac{\sWSRT}{\sNVSS} = \frac
    {1 \pm \left( C_1^2 + \left( \frac{C_2}{\sSNR} \right)^2\right)^\frac{1}{2}}
    {1 \pm \left( C_1^2 + \left( \frac{16.1 \times C_2}{\sSNR} \right)^2\right)^\frac{1}{2}}
\label{modelq3}
\end{equation}

\noindent based on $\sigNVSS = 16.1 \times \sigWSRT$, which implies
$\SNRWSRT = 16.1 \times \SNRNVSS$. The results using Eqn~\ref{modelq3}
has been overplotted on Fig.~\ref{fluxSigma} such that the maximum
(upper) envelope is given by setting the $\pm$ to $1+()/1-()$, and the
minimum envelope by $1-()/1+()$ in the equation. The $C$ parameters
have been set to $0.04$ and $1.3$ respectively, identical to the
values in \citet{rengelink97}. The model is most sensitive to the
$C_1$ value, which basically sets the envelope separation at high
\SNR. The $C_2$ value, which scales the \SNR\ dependence is far less
constrained. Values of $C_2=2$ (e.g., Kaper et al. 1966) are not
excluded. Given the assumptions and the assumed uncertainties about
the \NVSS\ errors, we adopt the \WENSS\ values of $0.04$ and $1.3$.
The quoted flux density errors in the final catalog are calculated
with these particular values.

\begin{figure}[tb]
\plotone{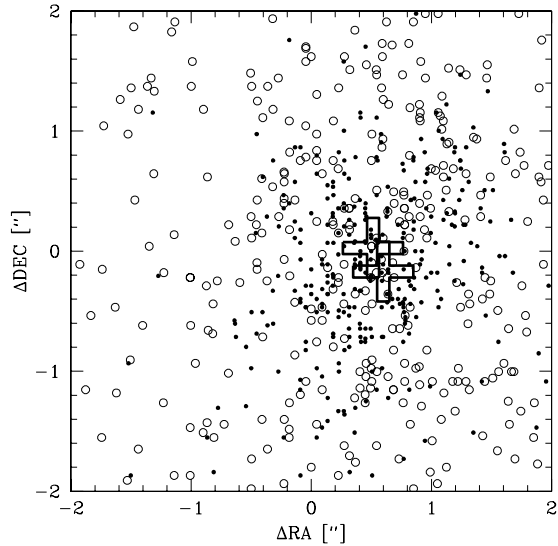}
\caption{\WSRT\ positional check against \FIRST\ and optical \APM\
positions. Only \WSRT\ point sources are used. \FIRST\ positions are
indicated by the solid circles, and \APM\ identifications are potted as
open circles. The larger scatter in the \APM\ correlation is due to the
association of radio positions with unrelated nearby optical
objects. The mean offsets (indicated by the crosses) are
$\Delta$RA$=+0.60\arcsec$, $\Delta$DEC$=-0.17\arcsec$ for \FIRST\, and
$\Delta$RA$=+0.51\arcsec$, $\Delta$DEC$=+0.02\arcsec$ for the \APM\
match.}
\label{posCheck}
\end{figure}

\begin{figure}[tb]
\plotone{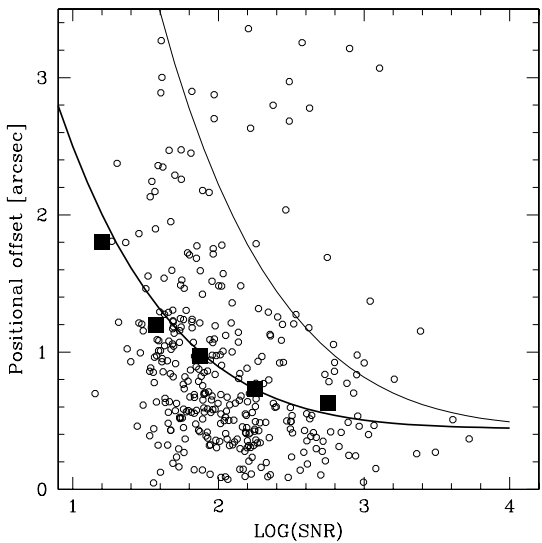}
\caption{Positional offsets from \FIRST\ / \APM\ positions as a
function of the signal-to-noise ratio (SNR). Overplotted as solid
squares are the 67 percentile values for a given SNR. The bottom curve
represents the best fitting 1$\sigma$ error envelope
(cf. Eqn.~\ref{modelpos}, with $C_1=0.44$ and $C_2=5.5$. The top curve
is identical to the one modeled for the \WENSS\ survey (Rengelink et
al. 1997).}
\label{posSNR}
\end{figure}

\subsection{Positional Accuracy} \label{posacc}

\begin{figure}[tb]
\plotone{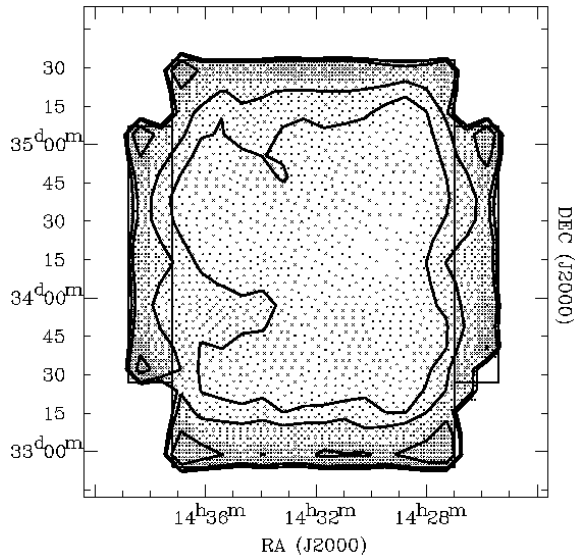}
\caption{Map of the rms noise inside the source extraction area
(outlined by the boxes). The noise levels and relative surface area
are, from inside out: $20-30$\mujy (60.9\%), $30-40$\mujy (23.9\%),
$40-80$\mujy (13.3\%), and $80-160$\mujy (2.0\%).}
\label{noiseMap}
\end{figure}

Optical identifications can only be securely made if the radio
positions are known accurately. The optical source density becomes
high enough towards fainter magnitudes to effectively have one
potential counterpart per beam. For instance, the mean source
separation in the Deeprange $I$-band field survey is about $17\arcsec$
at the 23$^{\rm rd}$ magnitude level \citep{postman98}. This
separation is actually smaller than the \WSRT\ beam size at 1400
MHz. Good positional matches are therefore essential.

We compared the cataloged positions for point sources against their
\FIRST\ positions and against Automatic Plate Measuring (\APM) machine
identifications. The \APM\ facility (in Cambridge, UK) catalogs
identifications and positions based on scanned UK and POSS~II Schmidt
plates, covering currently more than 15\,000 square degrees of sky.

The relative offsets for the individual sources are plotted in
Fig.~\ref{posCheck}. It is clear that the \FIRST\ and \APM\ positions
agree rather well with each other (indicated by the crosses), but that
our positions are systematically off in RA. Without more frequent
observations of additional calibrators, which would adversely affect
our uv coverage, astrometric accuracy of the \WSRT\ is known not to be
better than about $0.5\arcsec$ (e.g., Oort \& Windhorst 1985),
consistent with our offset value.  We corrected all the positions in
RA with $-0.56\arcsec$, i.e., the mean of the \FIRST\ and \APM\ RA
offsets. This correction corresponds to about 4\%\ of the beam width,
small but significant enough when accurate positional coincidences are
needed.

Analogously to Eqn.~\ref{modelq1} for the flux density errors, we can
describe the flux density dependence on positional accuracy in the form:

\begin{eqnarray}
\sigma_{\alpha,\delta} & = & \left( C^2_1 + C^2_2 \left(
\frac{\sigma_{\mbox{rms}}}{S} \right)^2 \right)^{\frac{1}{2}} \nonumber \\ 
& = & \left( C^2_1 + C^2_2 \left( \mbox{SNR} \right)^{-2} \right)^{\frac{1}{2}}
\label{modelpos}
\end{eqnarray}

The absolute distances from the \FIRST\ positions have been plotted in
Fig.~\ref{posSNR} as a function of signal-to-noise. Our survey
positions have been corrected for the RA offset first. Since the
\FIRST\ resolution is higher than our survey and may lead to 2 (or
more) \FIRST\ catalog positions for any of our positions, we only
considered \FIRST\ point sources within our survey field. The inclusion
of resolved sources (either in \FIRST\ or our survey) unnecessarily
complicates the comparison.

In Fig.~\ref{posSNR} a clear decrease in positional offset with
increasing \SNR\ can be seen. To characterize this trend, we fitted
Eqn.~\ref{modelpos} to the 67th percentile points (solid squares) in
order to get a 1$\sigma$ positional error estimate. The best fitting
values for the constants are: $C_1 = 0.44$ and $C_2 = 5.5$, both in
arcseconds. An outer envelope to the offset distribution is given by
$C_1 = 0.44$ and $C_2 = 15.0$. The value for $C_2$ is actually the
mean beam size (taken to be 20\arcsec) divided by 1.3, a value
identical to the one quoted for the \WENSS\ survey
\citep{rengelink97}. We adopt the first set of constants (the
1$\sigma$ equivalents) for our source catalog.

\subsection{Completeness and Reliability} \label{complreliab}

The background noise in our survey is not uniformly flat, but has a
marked upturn towards the edges. The tiling was set up in such a way
that in the interior regions the noise should be flat. This can be
verified in Fig.~\ref{noiseMap}, which plots the actual background
noise. The median noise level of the inner parts is 28$\mu$Jy.  The
large 30 to 40\mujy ``intrusion'' at 14$^{\rm h}$36$^{\rm m}$,
34$^{\rm d}$00$^{\rm m}$ is most likely due to the somewhat higher
noise levels in those four particular pointings.

\begin{figure*}[tb]
\epsscale{2.0} 
\plotone{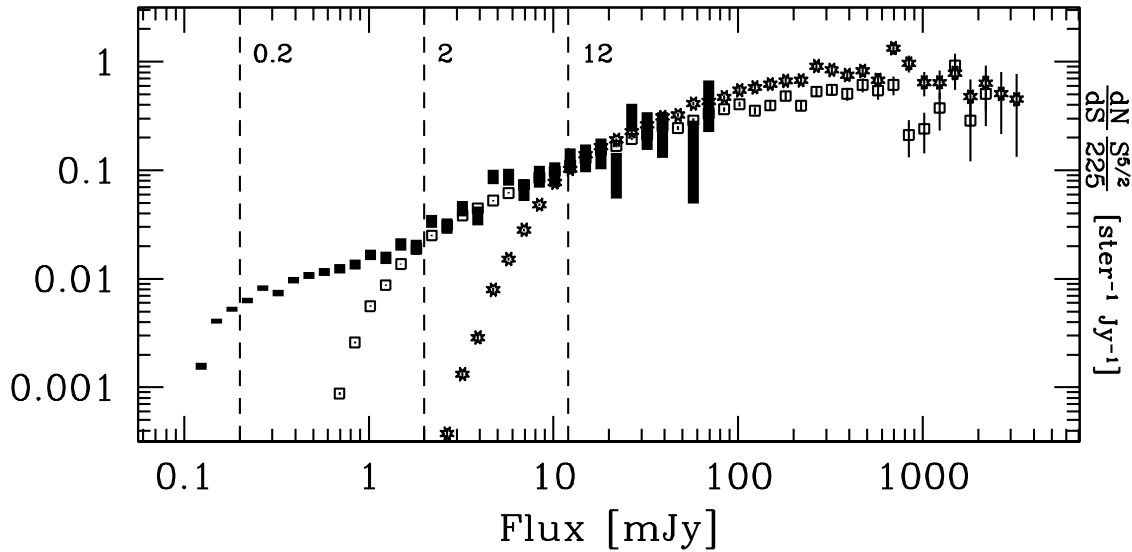}
\caption{Differential source counts as function of flux density $S$
(vertical bars). The counts have been normalized to the expected
number for a Euclidean universe. The lengths of the vertical bars are
set by the $\sqrt{N}$ error-bars. Overplotted are the differential
source counts for the \NVSS\ (stars) and \FIRST\ (open squares)
surveys. Note the onset of incompleteness in \NVSS\ at 12 mJy, in
\FIRST\ around 2 mJy, and in our survey at about 0.2 mJy.}
\label{dnds}
\end{figure*}

The survey completeness can be gleaned from Fig.~\ref{dnds}, in which
the differential source counts are plotted against flux.  The number
counts have been normalized by the expected number in a Euclidean
universe, given by $N_{\mbox{Eucl}} = K_\nu S_\nu^{-5/2}$, with the
unitless constant $K_\nu = 225$ (consistent with e.g., Oort \&
Windhorst 1985 and Oort 1987, but see \citet{wall94} who lists
$K_\nu=200$, however). Our source counts are compared to the ones
based on the \NVSS\ and \FIRST\ catalogs, which due to their much
larger survey area extend further towards higher flux densities. There
is good agreement (within the 1$\sigma$ errorbars) over the range 10
-- 100 mJy between the surveys. The small deviation in our
differential counts around the 4 mJy bin appears to be real and might
indicate the presence of an overdense region within our field (e.g., a
cluster). Since our survey field is relatively small at 6.68 square
degrees, any local overdensity could skew the number counts
significantly. The \NVSS\ and \FIRST\ number counts are not affected
by this, and serve as a useful baseline.

All three plotted surveys have different completeness limits, and if
one marks the first systematic deviation from a low-order polynomial
fit as the completeness limit, we measure 12 mJy for \NVSS, 2 mJy for
\FIRST, and 0.2 mJy for our survey. However, since the noise in our
survey is not constant across the source extraction area, but varies
within the 30\%\ level over $\sim$90\%\ of the survey, the 0.2 mJy
value is not strictly correct. It represents a mean value over the
survey area, with completeness levels slightly lower and higher 
in the inner and more outer parts of the survey respectively.

\subsection{Source Confusion}

With the $13\arcsec\times27\arcsec$ beam size and the faint flux
density levels reached in this survey, considerable source confusion
might be present. We therefore modeled this by randomly distributing
the cataloged source population over the survey area a large number of
times (10$^{4-5}$). Each time the likelihood of having close pairings
of sources was recorded. Given a large enough sampling, a more or less
accurate estimate of the frequency of occurrence is possible.  The
results are given in Table~3. Since we used the actual catalog, a
strong flux dependency is to be expected. In other words, having two
bright ($\ga10$ mJy) objects very close together almost always means
they are physically associated, whereas two faint ($\la0.5$ mJy) objects
with a similar separation are most likely unconnected. Based on the
numbers from Table~3 we can state that sources with a double
morphology, with flux densities $> 3$ mJy, and angular separations of
$<1$ arcminute have a 96\%\ chance of being true physical doubles.
All of the listed resolved sources with flux densities $> 10$ mJy in
Table~2 should therefore be considered single physical entities and
hence accurately classified.

Assuming the approximately 2-fold increase in unrelated object count
continues toward fainter flux levels, a similar \WSRT\ survey would
become confusion dominated (i.e. with on average 2 faint objects in
the synthesized beam) around the 15 $\mu$Jy mark. For this survey,
with a 5$\sigma$ limit of 140 $\mu$Jy, on average 10\%\ of the beams
are confused.

\subsection{Source Catalog}

The final catalog, which lists every source with a flux density over 5
times the local $\sigma$ (corresponding to 140\mujy\ in the center of
the survey field), contains 3172 sources. Roughly 10\%\ of these are
resolved (316) by the $13\arcsec\times27\arcsec$ beam. A complete
break down of source morphology is given in Table~4. More complete
catalogs with varying threshold $\sigma$'s are available from the ftp
site. The total number of included sources decreases with increasing
\SNR\ limits: 3172, 2767, 2367, 2061, 1854, and 1692 sources for 5, 6,
..., 10$\sigma$ thresholds respectively. Also, detailed radio maps of
the complete survey area are provided, with the cataloged sources
clearly indicated. This will allow for a direct visual assessment
whether a particular source is to be considered real or not.

The 73 brightest resolved objects have been listed in Table~2, all of
which have flux densities in excess of 10 mJy. Contour plots for these
particular sources are presented here in Fig.\ref{mos000}.

\begin{figure}[tb]
\epsscale{1.0}
\plotone{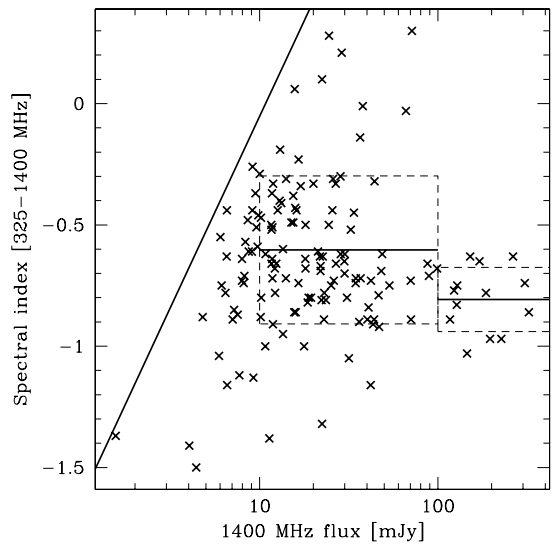}
\caption{Spectral index distribution as as a function of 1400MHz flux.
The 325MHz data is from the \WENSS\ survey. The slanted line
represents the \WENSS\ sensitivity limited spectral index, using a
3$\sigma$ detection threshold of 11 mJy at 325MHz. The two solid
horizontal lines are the spectral index means for the flux density
ranges 10 to 100, and 100 to 1000 mJy, respectively. The dashed boxes
outline the 1$\sigma$ area.}
\label{spixDist}
\end{figure}

Our survey catalog contains 143 sources which are also detected in the
\WENSS\ survey.  Fig.~\ref{spixDist} plots the 325$-$1400MHz spectral
index distribution of these sources. There appears to be a trend for
the more luminous sources ($> 100$ mJy) to have slightly steeper
spectral indices than the fainter part of the sample ($<100$ mJy). The
actual mean values are $-0.60\pm0.31$ and $-0.81\pm0.13$ for the flux
bins $10-100$ and $100-1000$ mJy respectively. The quoted errors are
the 1$\sigma$ standard deviations. This overall flattening of the
spectral index with decreasing flux density levels is consistent with
the data presented in Fig.~\ref{specPop}, which shows the change in
radio source population as a function of flux density based on the
\WENSS\ and \NVSS\ surveys. Unfortunately, we cannot use our much
deeper survey (compared to \NVSS) to extend this towards even lower
flux densities.  To the left of the slanted line in
Fig.~\ref{spixDist} the \WENSS\ survey was not deep enough to detect
the radio sources at 325MHz. We will use our deep VLA observations of
the Bootes field for this purpose.

Ten of the radio sources in our catalog have redshifts given in the
literature, and these are listed in Table~5. Alternative (radio
catalog) names for some of our objects are given in the footnotes to
Table~2.

\section{Summary}

We presented the results from our deep \WSRT\ observations of the
Bootes Deep Field. The survey reached a 1$\sigma$ limiting flux
density of 28\mujy\ in the central region, and 3172 sources were
detected above the 5$\sigma$ level in a 6.68 square degree area.  The
survey is deep enough to sample the change in radio source population
properties at the few mJy level. In combination with our lower
frequency \VLA\ data, these datasets will provide key information
pertaining, among other things, to the nature and evolution of radio
sources, both in the local and the high redshift universe.

\acknowledgments

The authors like to thank the referee, Dr. Fomalont, for useful
comments and suggestions. WDV and WVB's work was performed under the
auspices of the U.S. Department of Energy, National Nuclear Security
Administration by the University of California, Lawrence Livermore
National Laboratory under contract No. W-7405-Eng-48. WDV thanks the
Nederlandse Organisatie voor Wetenschappelijk Onderzoek (NWO) for
their generous allocation of his visitors grant to Dwingeloo, which
got this project underway. The Westerbork Synthesis Radio Telescope is
operated by the ASTRON (Netherlands Foundation for Research in
Astronomy) with support from the Netherlands Foundation for Scientific
Research (NWO).  This research has made use of the NASA/IPAC
extragalactic database (NED) which is operated by the Jet Propulsion
Laboratory, Caltech, under contract with the National Aeronautics and
Space Administration.

\clearpage 

\begin{figure*}
\epsscale{2.0}
\plotone{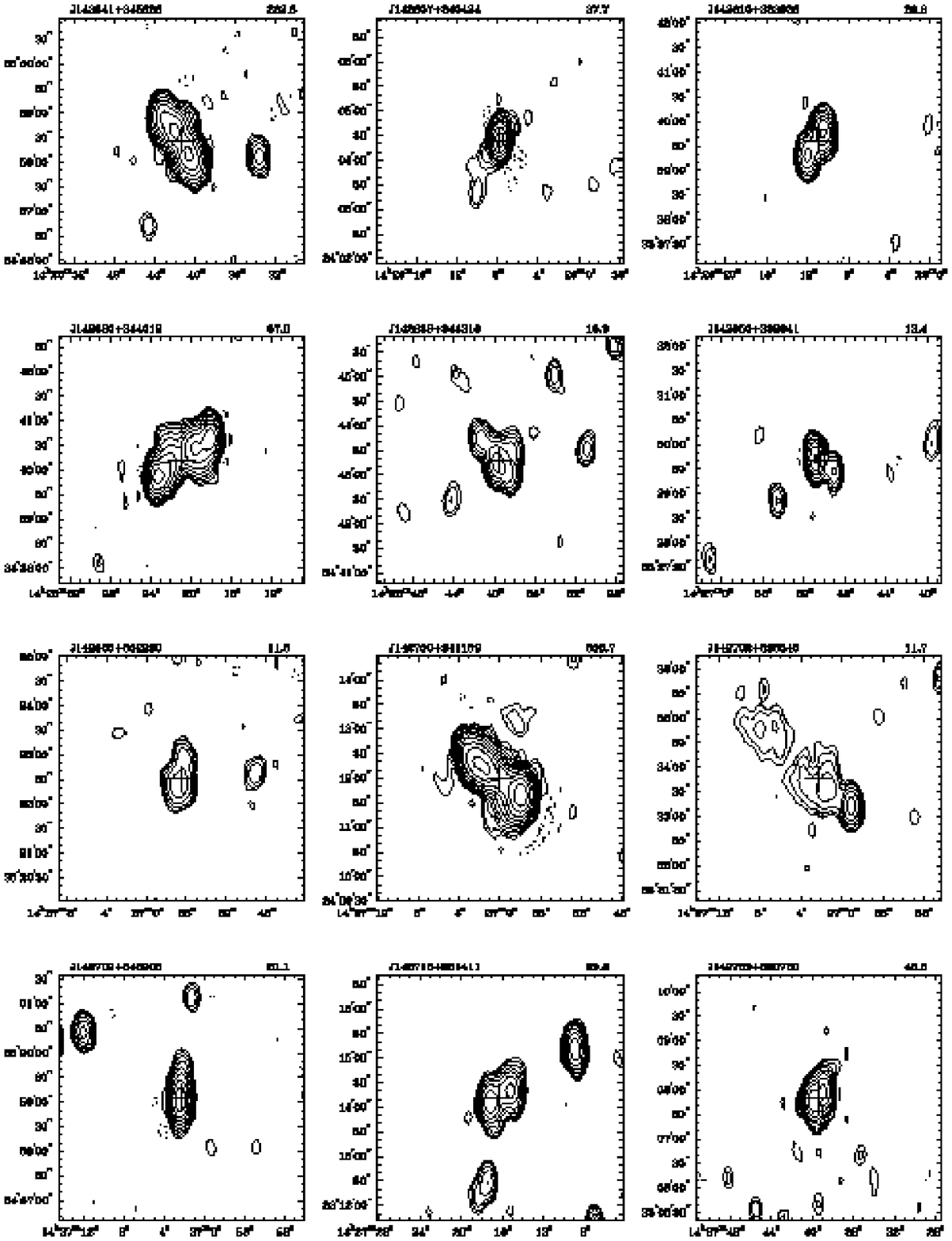}
\caption{Plots of resolved sources with $S>10$ mJy. The contour levels
are given by $(1+2^n)\sigma$, with $n=1,2,...$ and $\sigma$ as the
local background noise. Negative contours have the same spacing, and
are plotted as dotted lines. The cross represents the nominal source
center, and the object name and 1400 MHz flux densities are given in
the upper left and right corners.}
\label{mos000}
\end{figure*}

\clearpage

\begin{figure*}
\figurenum{11}
\epsscale{2.0}
\plotone{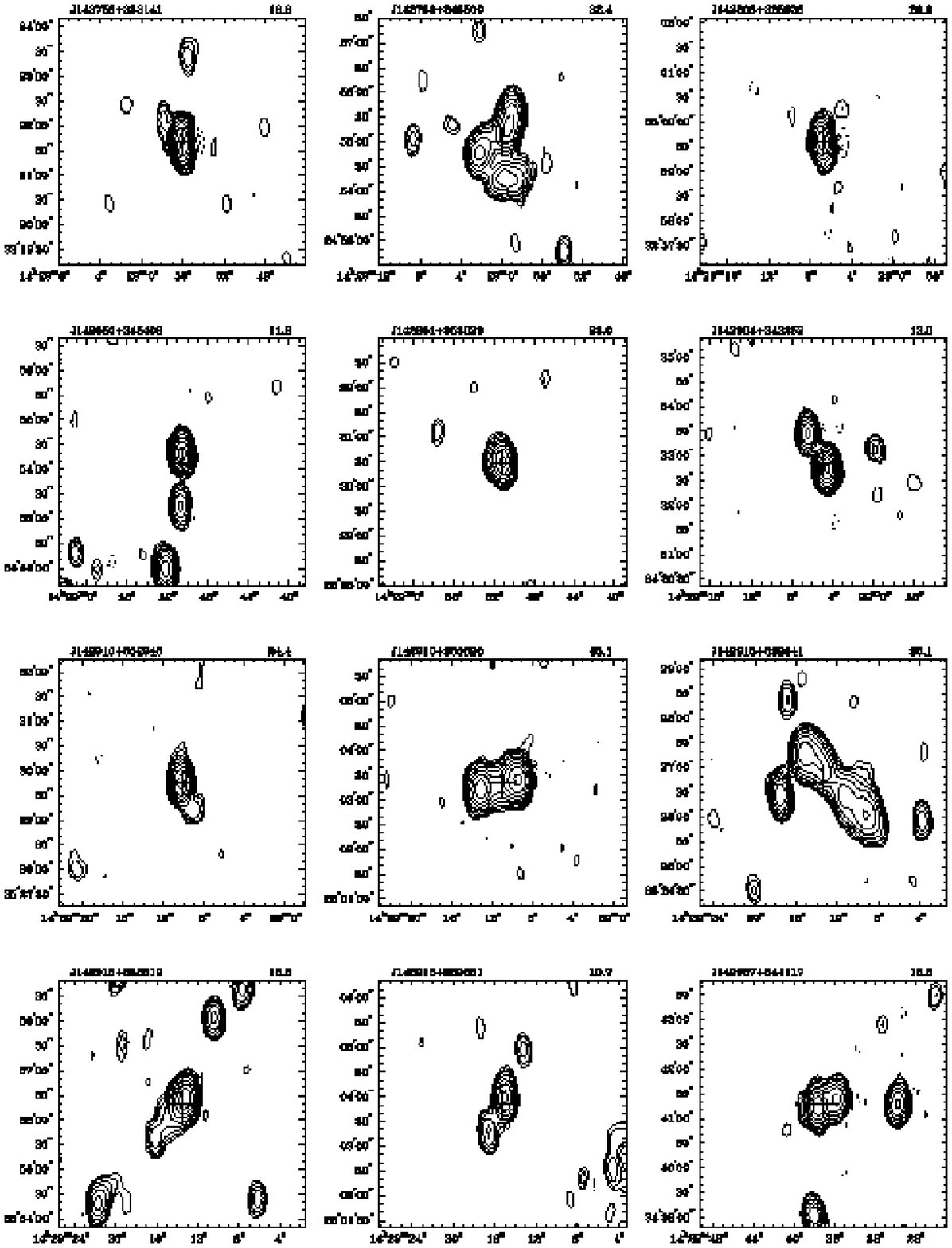}
\caption{Continued}
\label{mos001}
\end{figure*}

\clearpage

\begin{figure*}
\figurenum{11}
\epsscale{2.0}
\plotone{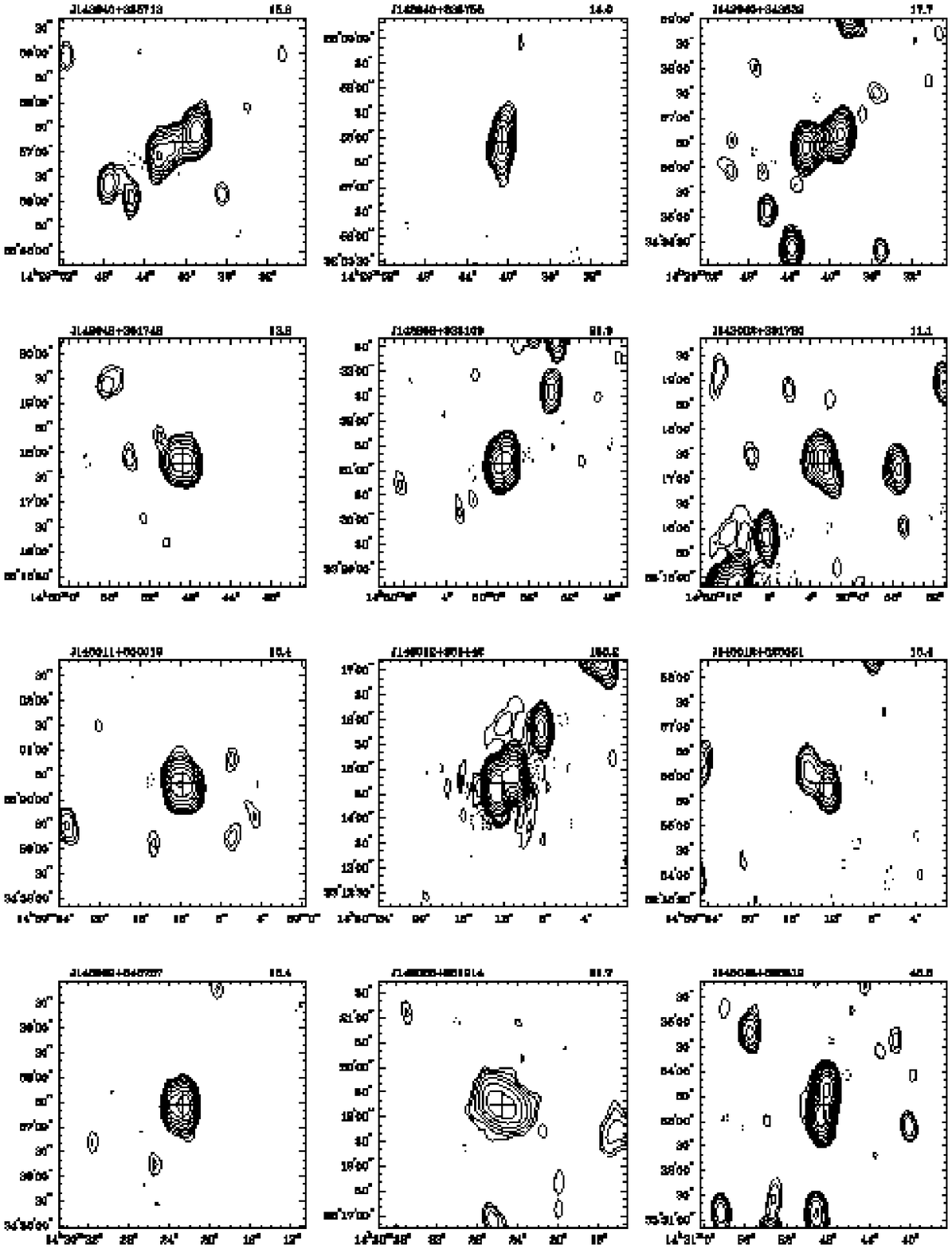}
\caption{Continued}
\label{mos002}
\end{figure*}

\clearpage

\begin{figure*}
\figurenum{11}
\epsscale{2.0}
\plotone{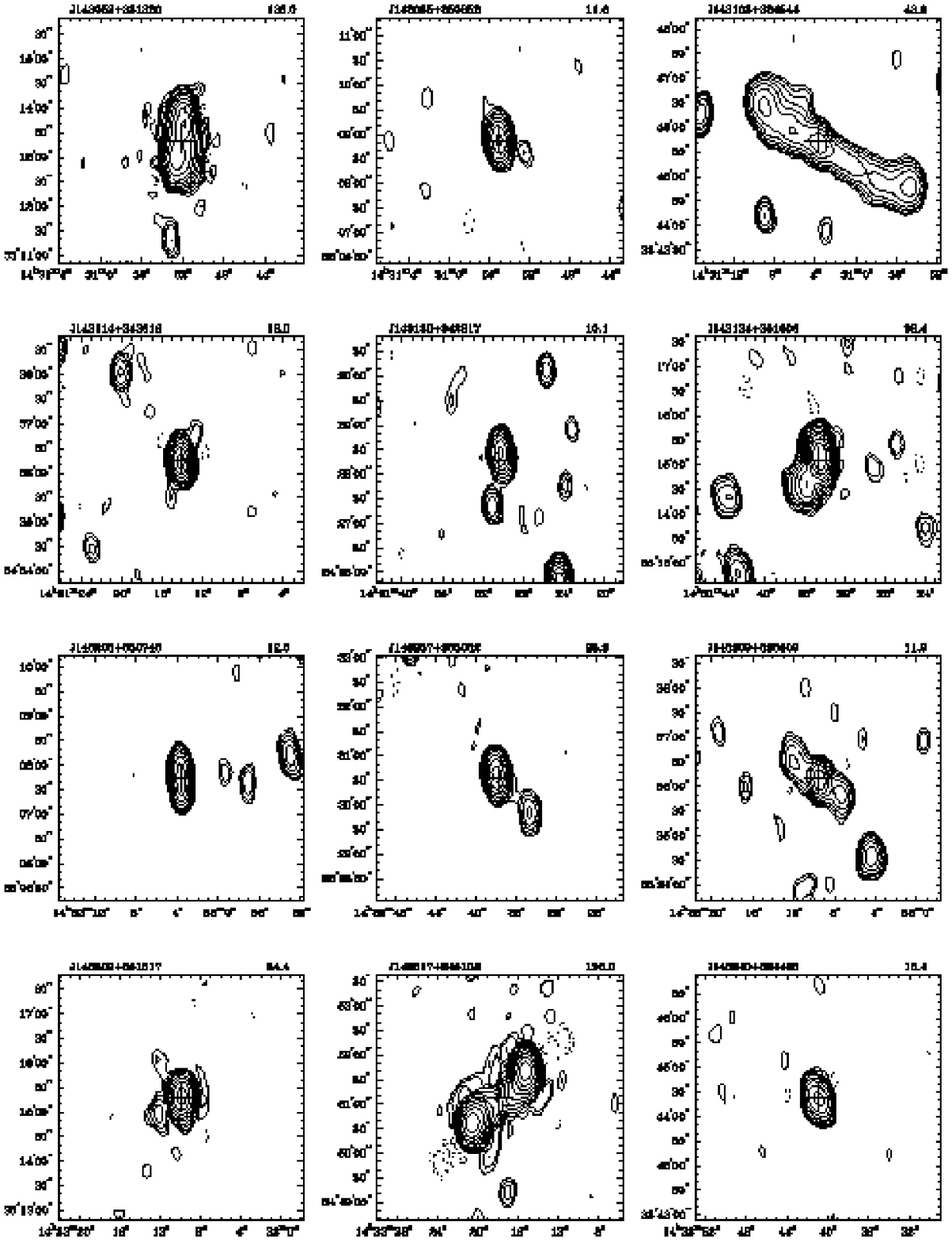}
\caption{Continued}
\label{mos003}
\end{figure*}

\clearpage

\begin{figure*}
\figurenum{11}
\epsscale{2.0}
\plotone{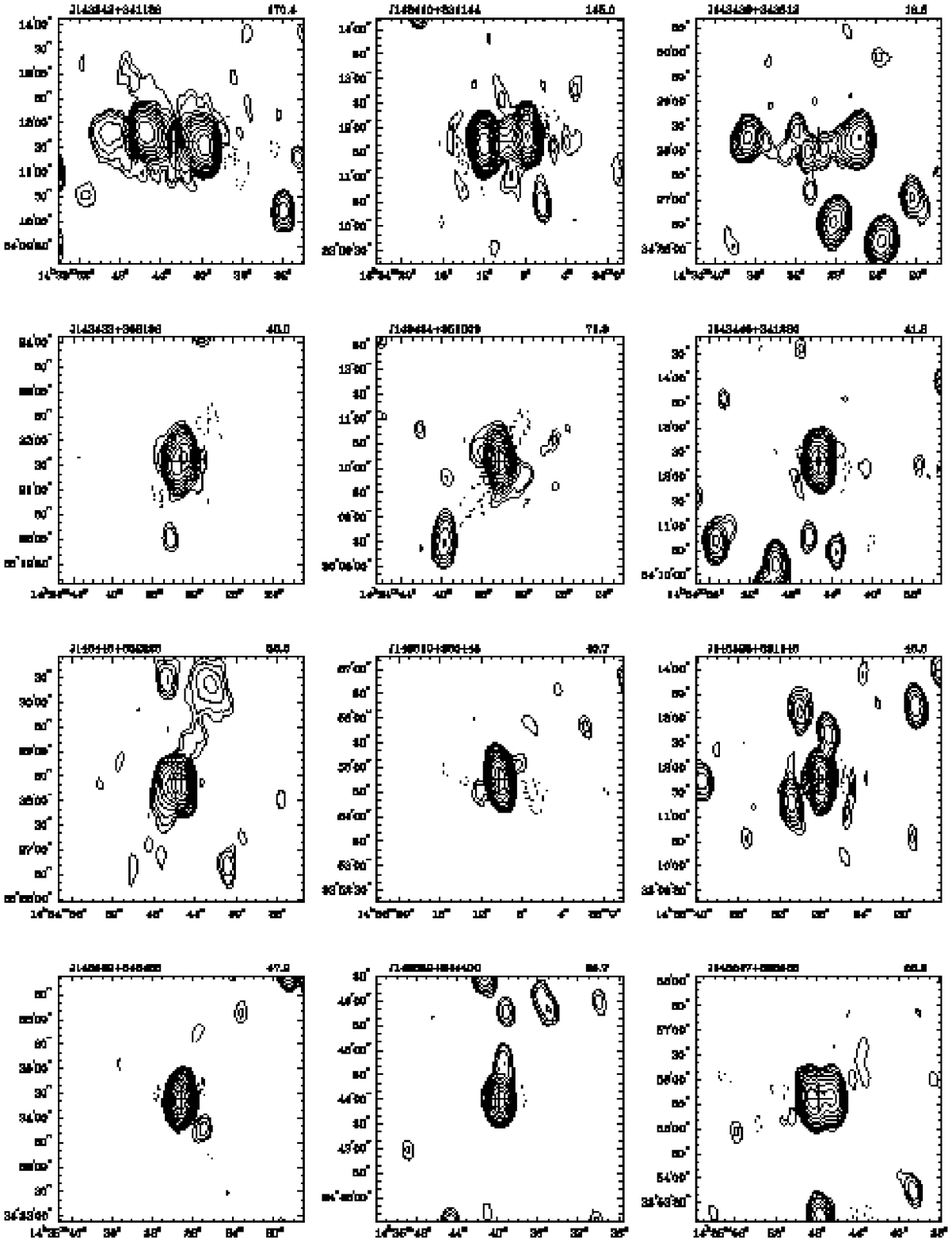}
\caption{Continued}
\label{mos004}
\end{figure*}

\clearpage

\begin{figure*}
\figurenum{11}
\epsscale{2.0}
\plotone{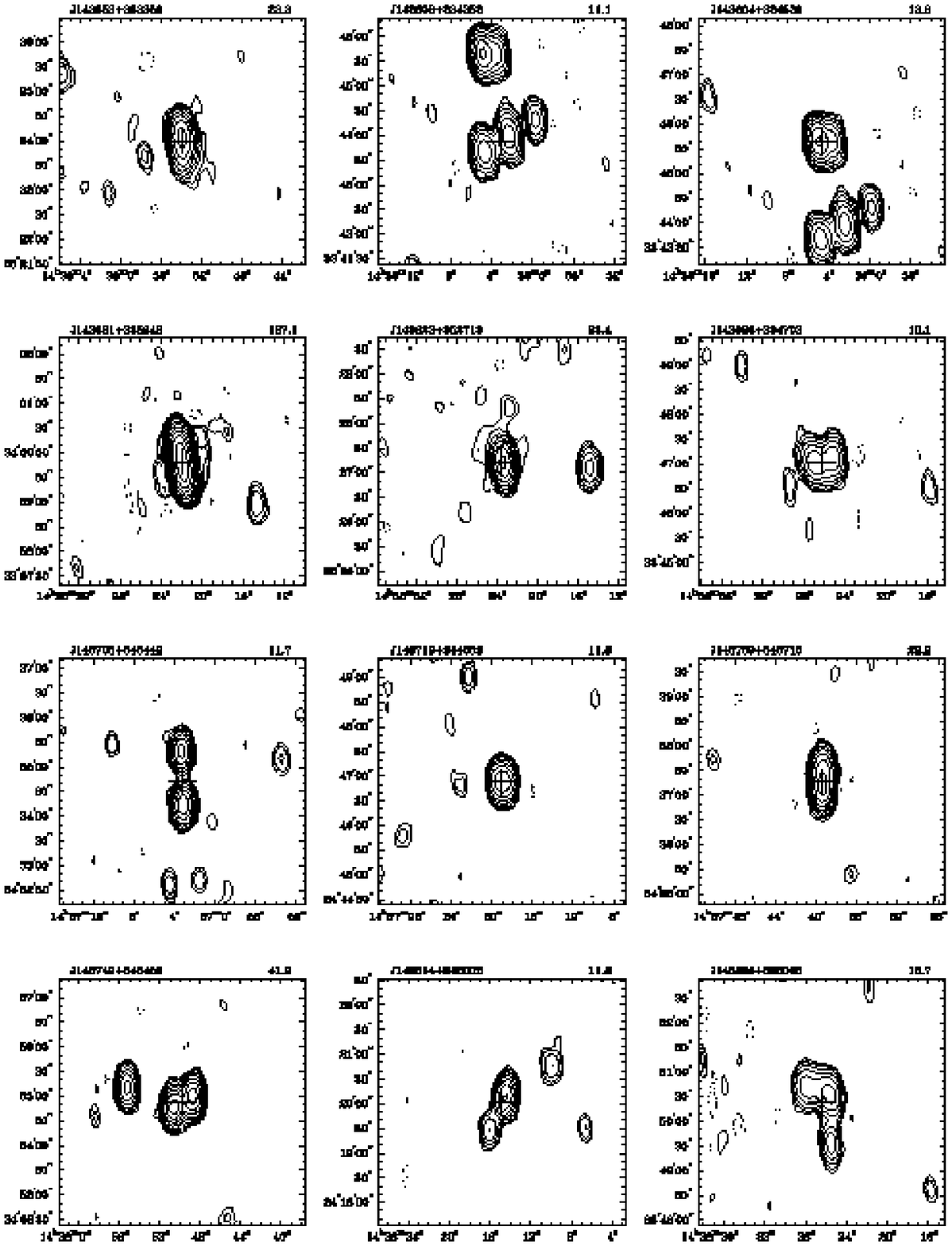}
\caption{Continued}
\label{mos005}
\end{figure*}

\clearpage

\begin{figure*}
\figurenum{11}
\epsscale{0.65}
\plotone{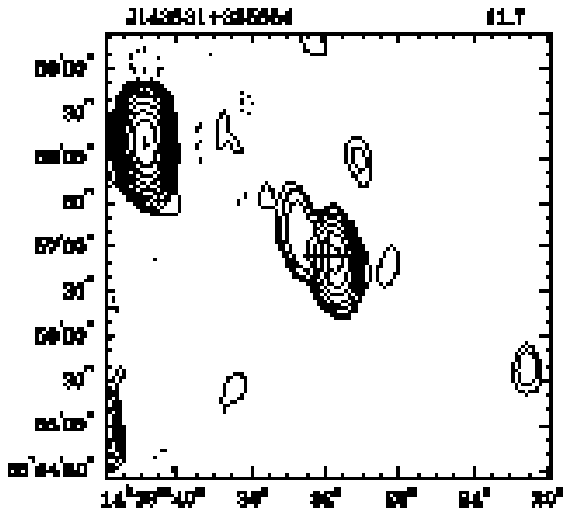}
\caption{Continued}
\label{mos006}
\end{figure*}

\begin{deluxetable}{lrccr}
\tablenum{1}
\tablecaption{Comparison between various radio surveys covering the Bootes field}
\label{surveyTab}
\tablehead{
  \colhead{Survey} & \colhead{Frequency} & \colhead{Resolution} & 
  \colhead{Flux density limit\tablenotemark{a}} & 
  \colhead{Detections in Bootes Field\tablenotemark{b}}
}
\startdata
WENSS       &  325 & $54\arcsec\times99\arcsec$   & 15  &  180 \\
VLA-Bootes  &  325 & $6\arcsec$                   & 0.5 &  1200\tablenotemark{c}\\
NVSS        & 1400 & $45\arcsec$                  & 2.5 &  438 \\
FIRST       & 1400 & $5\arcsec$                   & 1.0 &  749 \\
WSRT-Bootes & 1400 & $13\arcsec\times27\arcsec$   & 0.140 & 3172 \\
87GB        & 4850 & $222\arcsec\times198\arcsec$ & 18  &   22 \\
\enddata
\vspace{-2mm}
\tablenotetext{a}{5$\sigma$ detection limit. Units are in milli-Jansky's.}  
\tablenotetext{b}{The number of radio sources/components within a
circular aperture with a radius of 5400$\arcsec$, centered on $14^{\rm
h}32^{\rm m}05\fs75$, $34\arcdeg16\arcmin47\farcs5$. Note that this
number is both depending on flux limit and resolution.}
\tablenotetext{c}{Expected.}
\end{deluxetable}

\begin{deluxetable}{lccccrccccc}
\tablewidth{7.25in}
\tablenum{2}
\tablecaption{Resolved sources with flux densities $>$10 mJy.}
\label{ResCatalog}
\tablehead{
  \colhead{Source} & 
  \colhead{RA (J2000)} & \colhead{Dec (J2000)} & 
  \colhead{$\delta$POS} &
  \colhead{F\tablenotemark{a}} & 
  \colhead{F$_{\mbox{int}}$} & \colhead{rms\tablenotemark{b}} & 
  \colhead{$\Theta\tablenotemark{c}_{\mbox{ma}}$} &
  \colhead{$\Theta\tablenotemark{c}_{\mbox{mi}}$} &
  \colhead{PA} & \colhead{LAS\tablenotemark{d}}
  \\
  \colhead{} & \colhead{} & \colhead{} &
  \colhead{[$\arcsec$]} & \colhead{} &
  \colhead{[mJy]} & \colhead{[mJy]} & \colhead{[$\arcsec$]} &
  \colhead{[$\arcsec$]} & \colhead{[$\arcdeg$]} & 
  \colhead{[$\arcsec$]} 
}
\startdata
J142541+345826 & 14 25 41.40 & +34 58 26.6 & 0.4 & M & $ 262.60\pm 10.5$  & 0.102 & 244 &  89 & 27 & 243 \\
J142607+340424 & 14 26 07.73 & +34 04 24.0 & 0.4 & M & $  37.73\pm 1.51$  & 0.048 & 111 &  49 &164 & 108 \\
J142610+333936 & 14 26 10.88 & +33 39 36.6 & 0.4 & M & $  29.78\pm 1.19$  & 0.050 & 159 &  67 &155 & 157 \\
J142620+344012 & 14 26 20.94 & +34 40 12.6 & 0.4 & M & $  87.01\pm 3.48$  & 0.034 & 231 &  77 &130 & 229 \\
J142639+344318 & 14 26 39.42 & +34 43 18.6 & 0.4 & M & $  15.88\pm 0.64$  & 0.031 & 150 &  91 & 22 & 148 \\
J142650+332941 & 14 26 50.53 & +33 29 41.6 & 0.4 & M & $  13.41\pm 0.54$  & 0.038 & 103 &  60 & 22 & 99.4 \\
J142656+352230 & 14 26 56.47 & +35 22 30.9 & 0.4 & M & $  11.31\pm 0.46$  & 0.066 & 177 &  76 &172 & 175 \\
J142659+341159\tablenotemark{e} & 14 26 59.96 & +34 11 59.4 & 0.4 & E & $ 305.74\pm 12.2$  & 0.042 & 229 &  81 & 48 & 227 \\
J142702+333346 & 14 27 02.23 & +33 33 46.1 & 0.4 & E & $  11.73\pm 0.47$  & 0.034 & 243 &  73 & 46 & 241 \\
J142702+345905 & 14 27 02.24 & +34 59 05.2 & 0.4 & M & $  21.14\pm 0.85$  & 0.031 & 150 &  56 &179 & 148 \\
J142716+331411 & 14 27 16.20 & +33 14 11.5 & 0.4 & M & $  29.81\pm 1.19$  & 0.049 & 145 &  90 &140 & 142 \\
J142739+330750\tablenotemark{e} & 14 27 39.33 & +33 07 50.3 & 0.4 & M & $  43.62\pm 1.75$  & 0.049 & 128 &  71 &161 & 125 \\
J142756+332141 & 14 27 56.02 & +33 21 41.3 & 0.4 & M & $  18.77\pm 0.75$  & 0.027 & 106 &  52 &  7 & 103 \\
J142759+345500 & 14 27 59.82 & +34 55 00.7 & 0.4 & M & $  32.37\pm 1.30$  & 0.026 & 178 &  94 &167 & 176 \\
J142806+325936 & 14 28 06.69 & +32 59 36.7 & 0.4 & M & $  29.93\pm 1.20$  & 0.067 & 109 &  54 &178 & 106 \\
J142850+345408 & 14 28 50.51 & +34 54 08.6 & 0.4 & M & $  11.18\pm 0.45$  & 0.024 & 180 &  36 &178 & 178 \\
J142851+353029 & 14 28 51.07 & +35 30 29.5 & 0.4 & M & $  22.95\pm 0.92$  & 0.050 & 130 &  79 &  4 & 127 \\
J142904+343252 & 14 29 04.89 & +34 32 52.4 & 0.4 & M & $  12.98\pm 0.52$  & 0.018 & 142 &  43 & 24 & 139 \\
J142910+352945 & 14 29 10.13 & +35 29 45.1 & 0.4 & M & $  24.37\pm 0.98$  & 0.045 & 112 &  50 &  8 & 109 \\
J142910+350320 & 14 29 10.88 & +35 03 20.6 & 0.4 & M & $  43.10\pm 1.72$  & 0.025 & 159 &  80 &108 & 157 \\
J142913+332641 & 14 29 13.34 & +33 26 41.3 & 0.4 & M & $  36.06\pm 1.44$  & 0.027 & 215 & 123 & 57 & 213 \\
J142913+335619 & 14 29 13.51 & +33 56 19.6 & 0.4 & M & $  18.51\pm 0.74$  & 0.027 & 172 &  76 &151 & 170 \\
J142915+330351 & 14 29 15.18 & +33 03 51.4 & 0.4 & M & $  10.71\pm 0.43$  & 0.044 & 151 &  53 &160 & 149 \\
J142937+344117 & 14 29 37.07 & +34 41 17.9 & 0.4 & M & $  15.33\pm 0.61$  & 0.022 & 104 &  76 &131 & 100 \\
J142940+335713 & 14 29 40.29 & +33 57 13.0 & 0.4 & M & $  15.76\pm 0.63$  & 0.026 & 178 &  68 &127 & 176 \\
J142940+325755 & 14 29 40.48 & +32 57 55.6 & 0.4 & M & $  14.04\pm 0.57$  & 0.071 & 162 &  58 &175 & 160 \\
J142940+343632 & 14 29 40.61 & +34 36 32.0 & 0.4 & M & $  17.68\pm 0.71$  & 0.021 & 164 &  72 &116 & 162 \\
J142948+351748 & 14 29 48.72 & +35 17 48.3 & 0.4 & M & $  13.18\pm 0.53$  & 0.031 & 125 &  97 & 25 & 122 \\
J142958+333109 & 14 29 58.44 & +33 31 09.6 & 0.4 & M & $  21.92\pm 0.88$  & 0.027 & 130 &  83 &166 & 127 \\
J143002+331720 & 14 30 02.68 & +33 17 20.2 & 0.4 & M & $  11.07\pm 0.44$  & 0.032 & 141 &  79 & 17 & 138 \\
J143011+350019 & 14 30 11.80 & +35 00 19.4 & 0.4 & M & $  15.40\pm 0.62$  & 0.025 & 120 &  93 &  4 & 117 \\
J143012+331442 & 14 30 12.02 & +33 14 42.4 & 0.4 & E & $ 185.15\pm 7.41$  & 0.034 & 142 &  61 &148 & 139 \\
J143012+325551 & 14 30 12.80 & +32 55 51.6 & 0.4 & M & $  15.43\pm 0.63$  & 0.083 & 126 &  64 & 32 & 123 \\
J143022+343727 & 14 30 22.68 & +34 37 27.0 & 0.4 & M & $  16.44\pm 0.66$  & 0.023 &  99 &  74 &  3 & 95.2 \\
J143025+351914\tablenotemark{e} & 14 30 25.43 & +35 19 14.6 & 0.4 & M & $  21.68\pm 0.87$  & 0.033 & 172 & 130 & 42 & 170 \\
J143048+333319\tablenotemark{e} & 14 30 48.35 & +33 33 19.9 & 0.4 & M & $  46.61\pm 1.86$  & 0.027 & 183 &  54 &171 & 181 \\
J143052+331320\tablenotemark{e} & 14 30 52.10 & +33 13 20.5 & 0.4 & M & $ 127.98\pm 5.12$  & 0.034 & 222 &  80 &174 & 220 \\
J143054+350852 & 14 30 54.99 & +35 08 52.5 & 0.4 & M & $  11.61\pm 0.47$  & 0.028 & 104 &  59 & 10 & 100 \\
J143103+334544 & 14 31 03.64 & +33 45 44.1 & 0.4 & E & $  43.89\pm 1.76$  & 0.026 & 443 &  81 & 60 & 442 \\
J143114+343616 & 14 31 14.08 & +34 36 16.4 & 0.4 & M & $  18.03\pm 0.72$  & 0.023 & 109 &  50 &175 & 106 \\
J143130+342817 & 14 31 30.34 & +34 28 17.9 & 0.4 & M & $  10.09\pm 0.40$  & 0.024 & 169 &  41 &173 & 167 \\
J143134+351506\tablenotemark{e} & 14 31 34.68 & +35 15 06.7 & 0.4 & M & $  98.43\pm 3.94$  & 0.033 & 130 &  58 &165 & 127 \\
J143203+330743 & 14 32 03.59 & +33 07 43.7 & 0.4 & M & $  12.57\pm 0.51$  & 0.037 & 169 &  61 &  2 & 167 \\
J143237+353032 & 14 32 37.63 & +35 30 32.3 & 0.4 & M & $  28.30\pm 1.13$  & 0.056 & 136 &  54 & 32 & 133 \\
J143309+333609 & 14 33 09.55 & +33 36 09.8 & 0.4 & M & $  11.86\pm 0.48$  & 0.026 & 152 &  64 & 47 & 150 \\
J143309+351517 & 14 33 09.92 & +35 15 17.8 & 0.4 & E & $  34.40\pm 1.38$  & 0.033 & 102 &  68 &175 & 98.4 \\
J143317+345108\tablenotemark{e} & 14 33 17.83 & +34 51 08.1 & 0.4 & E & $ 195.03\pm 7.80$  & 0.035 & 244 &  56 &136 & 243 \\
J143340+334423 & 14 33 40.88 & +33 44 23.3 & 0.4 & M & $  15.39\pm 0.62$  & 0.029 & 128 &  80 & 17 & 125 \\
J143341+341138\tablenotemark{e} & 14 33 42.00 & +34 11 38.1 & 0.4 & E & $ 170.40\pm 6.82$  & 0.035 & 181 &  64 & 76 & 179 \\
J143410+331144 & 14 34 10.42 & +33 11 44.2 & 0.4 & M & $ 144.96\pm 5.80$  & 0.039 & 142 &  67 &101 & 139 \\
J143429+342812 & 14 34 29.69 & +34 28 12.2 & 0.4 & E & $  18.58\pm 0.74$  & 0.023 & 251 &  60 & 88 & 250 \\
J143433+352136 & 14 34 33.20 & +35 21 36.3 & 0.4 & M & $  39.99\pm 1.60$  & 0.040 & 113 &  67 &167 & 110 \\
J143434+351009 & 14 34 34.23 & +35 10 09.6 & 0.4 & M & $  71.27\pm 2.85$  & 0.036 &  96 &  53 &  6 & 92.1 \\
J143445+341220 & 14 34 45.20 & +34 12 20.3 & 0.4 & M & $  41.91\pm 1.68$  & 0.028 & 104 &  61 &  5 & 100 \\
J143445+332825 & 14 34 45.36 & +33 28 25.8 & 0.4 & M & $  36.54\pm 1.46$  & 0.027 & 157 &  47 &163 & 155 \\
J143510+335445 & 14 35 10.11 & +33 54 45.2 & 0.4 & M & $  40.66\pm 1.63$  & 0.034 & 110 &  55 &  5 & 107 \\
J143528+331145 & 14 35 28.15 & +33 11 45.5 & 0.4 & M & $  48.45\pm 1.94$  & 0.033 & 108 &  58 &158 & 105 \\
J143529+343423 & 14 35 29.15 & +34 34 23.1 & 0.4 & M & $  47.94\pm 1.92$  & 0.024 &  99 &  54 &  3 & 95.2 \\
J143539+344400 & 14 35 39.74 & +34 44 00.7 & 0.4 & M & $  26.71\pm 1.07$  & 0.024 & 109 &  50 &178 & 106 \\
J143547+335536 & 14 35 47.78 & +33 55 36.7 & 0.4 & M & $  53.21\pm 2.13$  & 0.034 &  99 &  93 & 62 & 95.2 \\
J143553+352359 & 14 35 53.88 & +35 23 59.8 & 0.4 & M & $  22.30\pm 0.89$  & 0.041 & 146 &  69 &  5 & 143 \\
J143602+334353 & 14 36 02.94 & +33 43 53.2 & 0.4 & M & $  11.15\pm 0.45$  & 0.030 & 210 &  92 &127 & 208 \\
J143604+334539 & 14 36 04.58 & +33 45 39.5 & 0.4 & M & $  13.84\pm 0.56$  & 0.031 & 100 &  83 & 12 & 96.3 \\
J143621+335949 & 14 36 21.94 & +33 59 49.5 & 0.4 & E & $ 127.08\pm 5.08$  & 0.037 & 163 &  64 &  5 & 161 \\
J143623+352713 & 14 36 23.49 & +35 27 13.7 & 0.4 & M & $  23.43\pm 0.94$  & 0.060 & 114 &  55 &179 & 111 \\
J143626+334703 & 14 36 26.19 & +33 47 03.1 & 0.4 & M & $  10.13\pm 0.41$  & 0.030 & 140 & 112 &117 & 137 \\
J143703+343442 & 14 37 03.20 & +34 34 42.0 & 0.4 & M & $  11.75\pm 0.47$  & 0.026 & 240 &  43 &  1 & 238 \\
J143718+344653 & 14 37 18.98 & +34 46 53.2 & 0.4 & M & $  11.76\pm 0.47$  & 0.030 & 126 &  85 &  2 & 123 \\
J143739+343716 & 14 37 39.43 & +34 37 16.5 & 0.4 & M & $  22.15\pm 0.89$  & 0.032 & 176 &  69 &179 & 174 \\
J143749+345452\tablenotemark{e} & 14 37 49.78 & +34 54 52.9 & 0.4 & M & $  41.87\pm 1.68$  & 0.045 & 132 &  89 &133 & 129 \\
J143814+342002 & 14 38 14.79 & +34 20 02.0 & 0.4 & M & $  11.83\pm 0.48$  & 0.056 & 145 &  52 &159 & 142 \\
J143826+335023 & 14 38 26.37 & +33 50 23.6 & 0.4 & M & $  15.66\pm 0.63$  & 0.056 & 235 & 103 & 15 & 233 \\
J143831+335654 & 14 38 31.83 & +33 56 54.5 & 0.4 & M & $  11.68\pm 0.47$  & 0.055 & 109 &  62 & 29 & 106 \\
\enddata
\tablenotetext{a}{Flag: S=point source, M=resolved, E=complex.}
\tablenotetext{b}{Local sky RMS (in mJy).}
\tablenotetext{c}{Apparent angular extent of 2.5$\sigma$
contour. Sources with $\Theta_{\rm maj}/\Theta_{\rm min} \approx 2$ and
PA$\approx 0\arcdeg$ are considered barely resolved.}
\tablenotetext{d}{Largest Angular Size. Resolved sources are
deconvolved with the beam size, point source sizes are approximated
by: ${\rm LAS} = \Theta_{\rm beam}\times \{0.04^2+(6.0/{\rm
SNR})^2\}^\frac{1}{4}$, cf.  Rengelink et al. (1997).}
\tablenotetext{e}{Alternative names: J142659+341159 = 7C 1412+344;
J142739+330750 = 7C 1425+333; J143025+351914 = NGC 5656;
J143048+333319 = 7C 1428+337; J143052+331320 = 7C 1428+334;
J143134+351506 = 7C 1429+354; J143317+345108 = 7C 1431+350;
J143342+341138 = 7C 1431+344; J143749+345452 = 7C 1435+351}
\end{deluxetable}

\begin{deluxetable}{clll}
\tablenum{3}
\tablewidth{4.4in}
\tablecaption{Source Confusion Limits.}
\label{confusion}
\tablehead{\colhead{Lower flux limit\quad} & 
    \multicolumn{3}{c}{Expected Number of Objects within radius R} \\
    \colhead{[mJy]} & \colhead{R = 30\arcsec\quad\quad\quad\quad} & 
    \colhead{R = 60\arcsec\quad\quad\quad\quad} & \colhead{R = 120\arcsec}}
\startdata
  0.05 & 0.11    &  0.43   &   1.70  \\
  0.10 & 0.10    &  0.43   &   1.69  \\
  0.20 & 0.074   &  0.29   &   1.18  \\
  0.40 & 0.041   &  0.17   &   0.65  \\
  0.80 & 0.025   &  0.097  &   0.38  \\
  1.60 & 0.012   &  0.057  &   0.23  \\
  3.20 & 0.0083  &  0.036  &   0.14  \\
  6.40 & 0.0052  &  0.019  &   0.073 \\   
 12.80 & 0.0026  &  0.0090 &   0.039 \\
 25.60 & 0.0012  &  0.0055 &   0.021 \\
 51.20 & 0.0007  &  0.0024 &   0.0095 \\
102.40 & 0.0002  &  0.0005 &   0.0031 \\
204.80 & 0.00002 &  0.0001 &   0.0007 \\
\enddata
\tablecomments{The listed counts are the number of unrelated objects
within a search radius R around a preselected target. For the total
source count within a radius R, 1 should be added to this count
therefore.  The lower 2 flux density bins are affected by the
incompleteness of the catalog at those levels; otherwise a factor of
$\sim2$ decrease in expected counts with increasing flux density
threshold seems to be present. Also note the surface area factor of 4
in count levels between the columns. Objects within the 30\arcsec\
radius are too close to be resolved by the \WSRT\ beam, and would
mistakenly be classified as a single source. To keep the number of
significant digits approximately constant we had to increase the
number of simulations with increasing flux density threshold.}
\end{deluxetable}

\begin{deluxetable}{lccr}
\tablenum{4}
\tablewidth{4.45in}
\tablecaption{Catalog morphology break-down.}
\label{cmorph}
\tablehead{\colhead{Morphology} & \colhead{Number} & \colhead{Density / sq.degree} &
\colhead{Example object}}
\startdata
Unresolved      & 2856 & 427.5 & \\
Barely resolved &   43 &   6.4 & J142851+353029 \\
Double          &  136 &  20.4 & J143703+343442 \\
Triple          &   13 &   1.9 & J143309+333609 \\
Asymmetric      &  112 &  16.7 & J143604+334539 \\
Complex / Other &   12 &   1.8 & J143429+342812 \\
\tableline
Total            & 3172 & 474.9 & \\
\enddata
\end{deluxetable}

\begin{deluxetable}{lllllrlr}
\tablenum{5}
\tablewidth{6.9in}
\tablecaption{Sources with known redshifts.}
\label{nedsources}
\tablehead{
  \colhead{Name} & \colhead{Alt. Name}  &
  \colhead{RA (J2000)\tablenotemark{a}} & \colhead{DEC\tablenotemark{a}} & 
  \colhead{ID} & \colhead{$z$} & \colhead{$V$} & \colhead{F$_{1400}$\tablenotemark{b}}
}
\startdata
J142744+333829 &                        &14 27 44.49 & +33 38 29.2 & G      & 1.237   & \nodata & 14.18 \\
J142823+331514 & UGC 09284              & 14 28 23.42 & +33 15 14.2 & G: Sa? & 0.01386 & 14.86   &  0.93 \\
J142932+333038 & CG 0447                & 14 29 32.66 & +33 30 38.4 & G      & 0.02638 & 17      &  0.12 \\
J142934+352742 & NGC 5646               & 14 29 34.07 & +35 27 42.2 & G: SBb & 0.02861 & 14.99   &  1.37 \\
J143025+351914 & NGC 5656               & 14 30 25.43 & +35 19 14.6 & G: SAab& 0.01051 & 12.73   & 15.33 \\
J143119+343803 & CG 0457                & 14 31 19.91 & +34 38 03.9 & G      & 0.01440 & 17.20   &  1.25 \\
J143125+331349 & CG 0459+0460           & 14 31 25.36 & +33 13 49.9 & G: S   & 0.02247 & 14.6    &  4.73 \\
J143156+333830 & VV 775                 & 14 31 56.15 & +33 38 30.1 & G: Irr & 0.03373 & 16      &  5.88 \\
J143232+340626 & LEDA 099838            & 14 32 32.42 & +34 06 26.3 & G      & 0.04264 & 17.96   &  0.21 \\
J143518+350709 & CG 0479                & 14 35 18.28 & +35 07 09.2 & G      & 0.02847 & 14.5    & 26.88 \\
\enddata
\tablenotetext{a}{Radio position.}
\tablenotetext{b}{Radio flux in mJy, with $\sigma=0.028$ mJy.}
\end{deluxetable}

\end{document}